# Potential-Programmed Operando Ensembles Govern Nitrate Electroreduction

Xue-Chun Jiang[1], Jia-Lan Chen[1], Wei-Xue Li*[1,2] and Jin-Xun Liu*[1,2]

[1]State Key Laboratory of Precision and Intelligent Chemistry, School of Chemistry and Materials Science, University of Science and Technology of China, Hefei, Anhui 230026, China

[2]Hefei National Laboratory, University of Science and Technology of China, Hefei 230088, China

Correspondence and requests for materials should be addressed to W.X.L. (Email: wxli70@ustc.edu.cn) or J.X.L. (Email: jxliu86@ustc.edu.cn)

## Abstract

Electrocatalyst surfaces continuously reorganize on the timescale of catalytic turnover, obscuring the identification of active sites under *operando* conditions and hindering rational catalyst design. Here, we resolve the *operando* Cu(111) electrolyte interface for nitrate-to-ammonia electroreduction ($NO_3RR$) via a multiscale modeling framework accelerated by a coverage-aware machine-learning potential. Rather than a single "average coverage" site, the working interface is a potential-gated statistical ensemble of 34 interconverting adsorbate motifs between –0.10 and –1.00 V (vs. SHE). Potential-driven shifts in motif populations produce a volcano-shaped activity trend peaking at –0.70 V, where the site-normalized turnover frequency reaches 0.015 $s^{-1}$ with nearly 100% Faradaic efficiency to ammonia. The activation barriers across >150 transition states collapse into a single linear relationship with the excess charge on the reacting Cu atoms ($\Delta q_{Cu}$), identifying interfacial charge redistribution as a unifying kinetic descriptor. The maximum activity arises not from uniform moderate coverage but from a $2NO/2NH_2$ quadrilateral microensemble that tunes $\Delta q_{Cu}$ to an intermediate value, simultaneously lowering the N–O cleavage and N–H formation barriers. Reconceptualizing "coverage" as an ensemble of local microenvironments decouples thermodynamic stability from catalytic productivity. This perspective furnishes a parameter-free strategy by controlling motif populations and interfacial charge via the potential to program high-coverage electrocatalysis beyond the $NO_3RR$.

## Keywords

# Introduction

Electrocatalysis has emerged as a pivotal technology for the net-zero chemical industry, converting renewable electrons into tailored chemical bonds under ambient conditions.[1-3] In addition to delivering green hydrogen and $CO_2$-derived fuels, this platform can also remediate environmental pollutants in real time, effectively weaving water treatment into chemical manufacturing.[4-6] The nitrate electroreduction reaction ($NO_3RR$) to ammonia exemplifies this dual benefit: it simultaneously purifies water and produces decentralized fertilizer,[7,8] potentially offsetting the traditional Haber–Bosch process, which consumes ~2% of global energy and emits over 400 Mt of $CO_2$ annually.[9] A decentralized electrolyzer that couples nitrate removal with renewable $NH_3$ synthesis could thus address both water security and the nitrogen footprint simultaneously.[10,11] However, achieving the high current densities ($\geq 1000$ mA cm$^{-2}$) required for industrial applications exposes a profound mechanistic blind spot.[12,13] Steep interfacial electric fields, fluxional intermediates, and a constantly reordering solvation shell continuously sculpt and resculpt the catalyst surface, obscuring the identification of *operando* active sites and hindering rational catalyst design.[14,15]

Copper stands out among earth-abundant cathodes for $NO_3RR$, maintaining >95% Faradaic efficiency (FE) across a wide potential window (–0.10 to –1.00 V vs. SHE) for the eight-electron, nine-proton reduction of nitrate to ammonia.[16-18] *Operando* microscopy and spectroscopy, however, depict a Cu(111) interface in perpetual motion, where coverage-dependent site adsorption increasingly reshapes the ensemble of reactive motifs.[19-23] Conventional mean-field microkinetic treatments that collapse this heterogeneity into a uniform "mean" site suppress motif distribution, interfacial electric fields and lateral coupling and consequently mispredict both electrocatalytic activity and selectivity.[24,25] Against this backdrop, three key questions define the frontier of $NO_3RR$ electrocatalyst design: (i) how the applied potential remodels the Cu surface coverage and redistributes adsorbate motifs under steady-state reaction conditions; (ii) how lateral interactions within high-coverage layers cooperate or compete to steer the activity and selectivity of the $NO_3RR$; and (iii) whether a single electronic descriptor can distill this configurational turmoil into quantitative, potential-dependent trends in reactivity. These issues extend beyond the $NO_3RR$ to other high-flux electrochemical

transformations—from $CO_2$ reduction ($CO_2$RR) to oxygen reduction (ORR)—where surface crowding and fluxionality dominate.[26-28]

Here, we introduce an *operando*-accurate multiscale framework that reconceives "coverage" as a population of local microenvironmental motifs with distinct kinetics. A coverage-aware machine-learning potential supplies DFT-level energetics nearly instantaneously, enabling exhaustive searches of the Cu(111) surface structure as a joint function of potential and coverage. From >450 low-energy configurations, we distinguish 34 recurrent adsorbate motifs spanning −0.10 to −1.00 V (vs SHE), each within ~0.50 eV of the global minimum and stabilized by cooperative interactions. Embedding this motif library into a grand-canonical, constant-potential, coverage-explicit microkinetic model yields a pronounced activity volcano peaking at −0.70 V, where the site-normalized turnover frequency reaches 0.015 $s^{-1}$ and nearly 100% $NH_3$ selectivity. Strikingly, the activation barriers across all the motifs collapse into a single linear relationship with the excess charge on the reactive Cu atoms ($\Delta q_{Cu}$), elevating the interfacial charge redistribution to a universal kinetic descriptor that links ångström-scale fluxionality to macroscopic rates. The peak activity does not arise from uniform "moderate coverage" but from a specific 2NO/2$NH_2$ quadrilateral microensemble that tunes $\Delta q_{Cu}$ to an intermediate value, simultaneously lowering the N–O scission and N–H formation activation barriers. By reconceptualizing "coverage" as an ensemble of local motifs and engineering local charge landscapes through potential control, this ensemble-based strategy offers a general, parameter-free blueprint for programming high-coverage electrocatalysts beyond the $NO_3$RR.

## Results

**Workflow for determination of the electrochemical surface states**

We established a closed-loop, three-stage workflow to resolve the working Cu(111) electrolyte interface under the $NO_3$RR (**Figure 1** and **Supporting Note 1**). In Stage I, we built an exhaustive grand-canonical free-energy library $\Delta G(U, \theta)$ for all intermediates and coverages. This library serves two purposes: it (i) provides on-the-fly energetics to a constant-potential microkinetic solver to deliver rates and self-consistent coverages and (ii) trains a reactive neural-network potential (NNP) that accelerates coverage-resolved structure searches. The NNP, fitted to >90,000 GC-DFT data points (RMSE < 2.10 meV atom$^{-1}$) (**Figure S1** and **S2**), evaluates candidate configurations generated by a genetic algorithm (GA) at each applied potential (from 0.00 V to −1.00 V in 0.10 V steps) and the coverage seeded from the preceding kinetic applied solution (**Figure S3**). Iterative crossover–mutation–selection yields >300 low-energy structures per potential. Hierarchical clustering compresses these into 34 Boltzmann-weighted motifs, each within ≤0.50 eV of the global minimum over −0.10 to −1.00 V (vs SHE), collectively defining the instantaneous *operando* ensemble. As a fidelity check, the NNP reproduces hallmark in situ features—such as trimeric *NO assemblies—while rapidly surveying potential- and coverage-dependent configurations.[29]

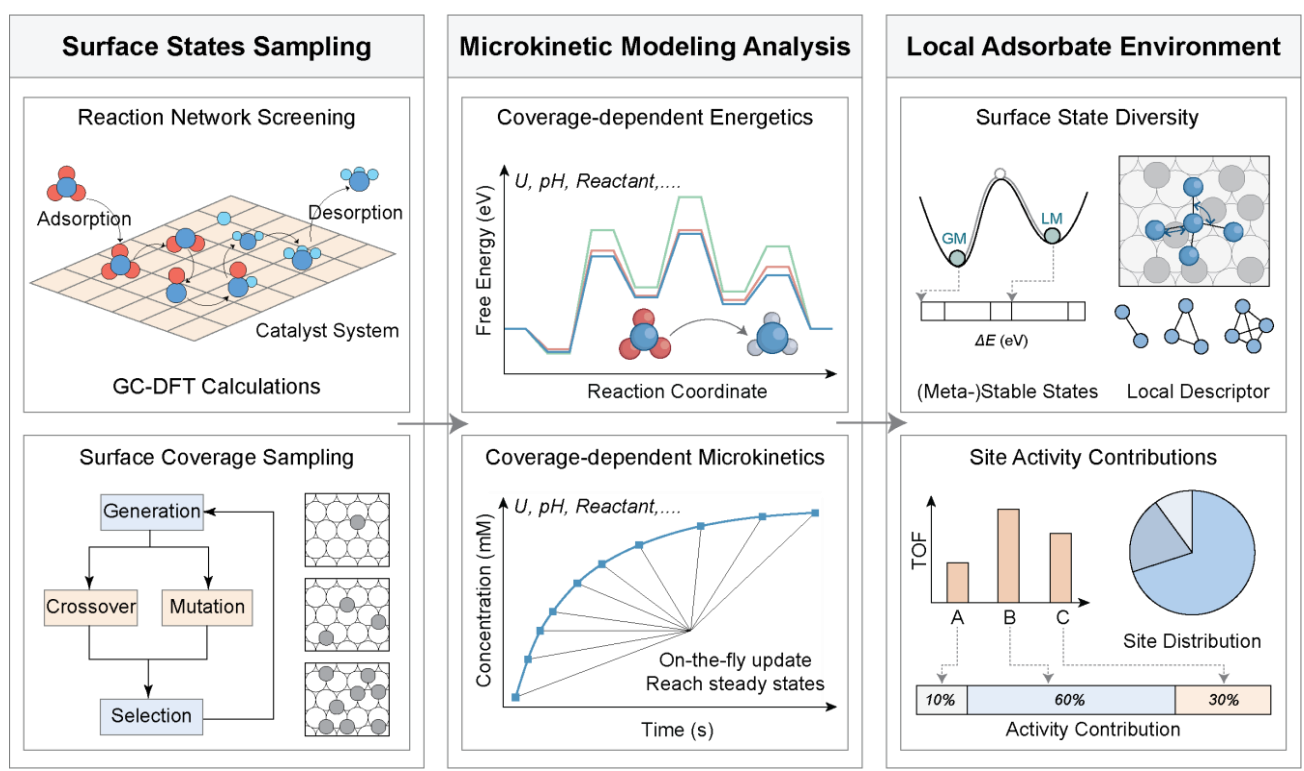

**Figure 1. Multiscale modeling framework for nitrate reduction on Cu(111).** The workflow integrates (i) reaction network generation and surface state sampling; (ii) coverage-dependent grand canonical DFT energetics coupled with constant-potential microkinetic modeling; and (iii) motif classification with quantification of each site's activity contribution. It includes surface state sampling (top-left) for determining surface interactions, microkinetic modeling analysis (top-center) to assess reaction pathways and free energy profiles, and local adsorbate environment characterization (top-right) to capture the influence of atomic-scale configurations on reaction behavior. The bottom

row details the computational methodologies employed: evolutionary algorithms for generating, mutating, and selecting surface configurations (bottom-left), time-dependent concentration profiles for dynamic simulations (bottom-center), and statistical analysis of adsorption states, illustrating the distributions of surface sites and reactant concentrations (bottom-right). This comprehensive modeling approach allows for a thorough understanding of reaction mechanisms and provides predictive insights into the efficiency of nitrate reduction processes on Cu(111).

In Stage II, kinetics are integrated "on the fly". For every motif, activation barriers ($E_a$) are recomputed with grand-canonical DFT, which explicitly captures the influence of the interfacial field, lateral interactions and site blocking as the surface evolves (**Supporting Methodology** and **Supporting Note 1**). Solving the resulting stiff ODE system to steady state yields turnover rates, product selectivities and degree-of-rate-control (DRC) coefficients under each condition. The microkinetic coverages are then fed back to seed a fresh GA search at the same potential, closing the loop (**Supporting Note 2**). After a few GA↔MKM cycles, the procedure converges to a self-consistent surface state that provides a smooth, potential dependent, Boltzmann-weighted ensemble of distinct configurations and a quantitative accounting of each motif's contribution to the overall $NO_3RR$ flux (**Supporting Note 3**). Therefore, rather than imposing a single uniform-coverage motif, we can enumerate and quantify every kinetically relevant adsorbate environment statistically on Cu(111) surface under *operando* conditions.

Ultimately, because every configuration is labeled by its local microenvironment, we deploy supervised classifiers trained on compact geometric and electronic descriptors to rank motif contributions to the overall activity and selectivity in Stage III. This analysis isolates the small set of motifs that disproportionately govern performance and distils governing principles that connect fluxional surface chemistry to macroscopic kinetics. The workflow is predictive and transferable in that it converts an *operando* snapshot—blurred by dynamic coverage and field effects—into a well-posed kinetic landscape without empirical tuning while resolving, atom-by-atom, how the $NO_3RR$ proceeds on Cu(111). The three stages of the approach are schematized in **Figure 1** and expanded upon in the following section. The same strategy generalizes to other high-coverage electrocatalytic transformations where population dynamics, not a single static structure, define the *operando* working active sites (**Figure S4**).

**Potential-dependent energetic landscape for NO$_3$RR on Cu(111)**

The nitrate-to-ammonia pathway on Cu(111) follows the canonical eight-electron sequence *NO$_3$ →*NO$_2$ →*NO →*NHO →*NHOH →*NH →*NH$_2$ →*NH$_3$, with surface-bound hydrogen (*H) serving as a quasimobile proton source (**Figure 2a** and **S5**).[30] Fast *H diffusion and cation–*NO$_x$ stabilization bias proton delivery to the surface and accelerate proton-coupled electron transfer (PCET) steps (**Figure S6** and **Supporting Note 4**).[31] To capture lateral interactions at high coverage (Stage I), we coupled a constrained genetic algorithm (Hookean constraints) to a trained neural network potential and enumerated >500 mixed-intermediate adlayers containing all relevant *NO$_x$/*NH$_x$ species (**Figure S7, S8** and **S9**).[32, 33]

To disentangle coverage from potential, we first fix the bias at 0.00 V and read out purely lateral effects (**Figure 2b**). Grand-canonical DFT reproduces established binding motifs across coverages: *NO$_3$ and *NO$_2$ bind bidentate via O, whereas *NO binds atop via N. Raising the total coverage from 1/6 to 1/2 ML preserves the *NO$_2$ geometry but drives *NO$_3$ from purely bidentate to mixed top+bridge—direct evidence of anion–anion repulsion (**Figure 2c** and **2d**). *NHO and *NHOH bind bidentate with O atop and N at a neighboring bridge (**Figure S10**). Downstream, *NH sits in hollows, *NH$_2$ at bridges, and *NH$_3$ atop (**Figure S8**).

Because *NO$_3$ and *NO are the dominant *operando* intermediates,[34,35] we quantified how precoverage by these species reshaped early thermodynamics and barriers (**Figure 2f**). NO$_3^-$ adsorption and the first N–O scission are strongly sensitive to crowding at 0.00 V: as the adlayer densifies (to ~4/5 ML), steric/electrostatic repulsion inverts the thermodynamics of *NO$_3$ ($\Delta G_{ads}$(NO$_3^-$) from −0.16 to +0.29 eV) and increases the step-specific $E_a$ (e.g., 0.72 → 0.99 eV) (**Figure 2b** and **2f)**. In contrast, *NO$_2$ cleavage results in a "Goldilocks" window: at intermediate $\theta_{NO3}$ coverage (~2/5 ML), the $E_a$ is minimized (0.18–0.31 eV). At low coverage, insufficient coadsorbate stabilization penalizes the transition state (TS); at high coverage, crowding hinders both adsorption and transition state (TS) formation for *NO$_2$ cleavage. This sharp contrast—*NO$_3$ activation worsens monotonically with crowding, *NO₂ cleavage optimized at intermediate coverage—already foreshadows the nonmonotonic kinetics under applied potentials.

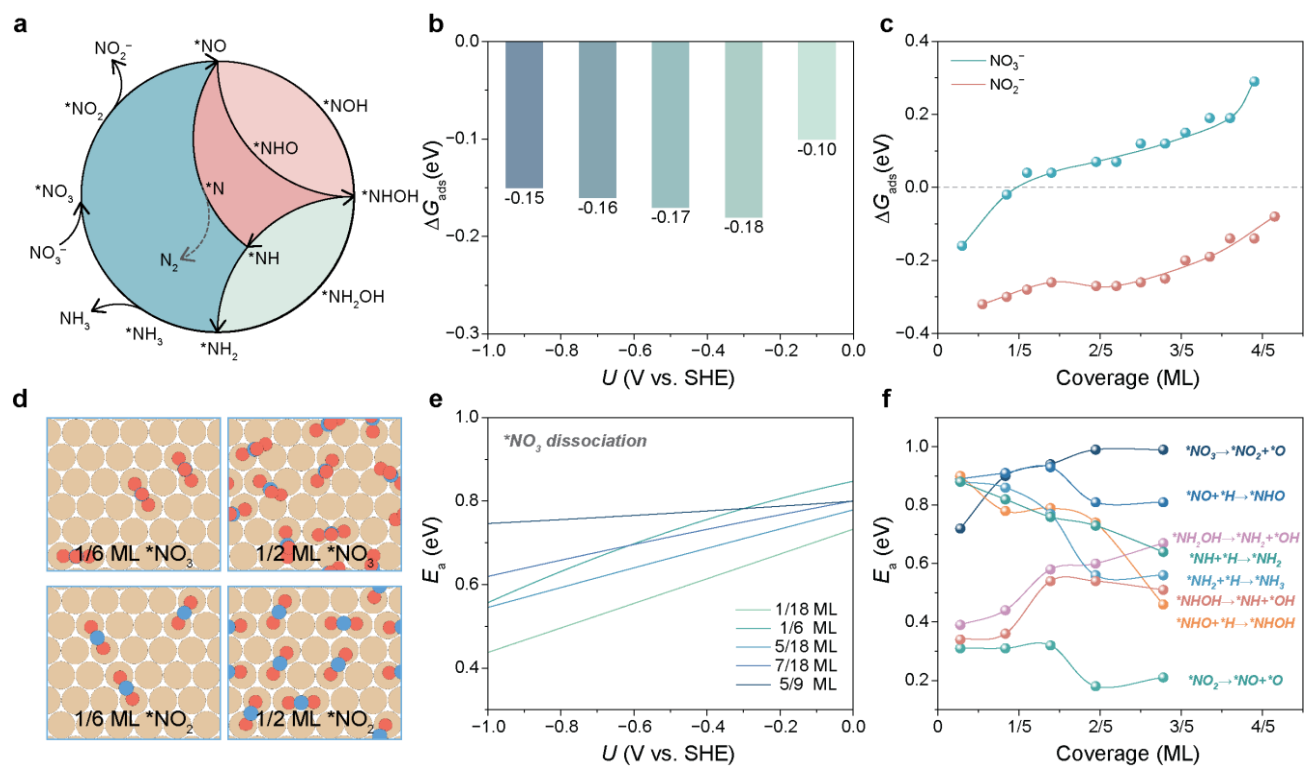

**Figure 2. Potential-driven surface coverage effect of the NO₃RR on Cu(111). (a)** Reaction network of the NO₃RR to NH₃ on Cu(111), with surface hydrogen (*H) as the proton source and NO₂⁻ as a byproduct. **(b)** Adsorption free energy ($\Delta G_{ads}$) of NO₃⁻ as a function of the applied potential (vs. SHE) at low coverage (1/6 ML). **(c)** Adsorption free energy ($\Delta G_{ads}$) for NO₃⁻ (green markers) and NO₂⁻ (pink markers) under varying coverages. **(d)** Adsorption configurations of *NO₃ and *NO₂ at low (1/6 ML) and high (1/2 ML) surface coverages. **(e)** Coverage-dependent activation energies ($E_a$) of *NO₃ dissociation, with solid lines corresponding to increasing *NO₃ coverages (from 1/18 to 5/9 ML). **(f)** Coverage-dependent activation energies ($E_a$) for each elementary step at 0.00 V vs. SHE.

The mid-sequence PCET steps respond in the opposite direction, creating a built-in tension. For *NO hydrogenation to *NHO, $E_a$ initially increases with coverage (0.89 → 0.93 eV from 1/18 → 5/18 ML) as local dipoles destabilize the PCET TS and then decreases under heavy crowding (0.62 eV at 5/9 ML) as short-range proton–relay geometries emerge (**Figure 2f**). *NHO hydrogenation to *NHOH shows a coverage minimum (0.46 eV near $\theta_{NO3}$ = 3/5 ML), whereas *NH hydrogenation to *NH₂ is comparatively insensitive. As *NHOH accumulates, hydrogenation becomes easier (1.04 → 0.71 eV), but *NH$_2$OH cleavage hardens (0.39 → 0.67 eV) because it blocks undercoordinated Cu. Because $E_a$(*NHOH scission) stays lower than $E_a$(*NHOH hydrogenation) and $E_a$(*NH$_2$OH cleavage), the network preferentially channels *NHOH to *NH + *OH rather than forming strongly bound *NH$_2$OH (**Figure S12–S14**).

Potential–coverage coupling creates nonmonotonic kinetics. At −0.60 V, the *NO$_3$ dissociation $E_a$ increases sharply with *NO$_3$ coverage—0.45 → 0.78 eV from 1/18 → 5/9 ML. However, at −0.40 V, it becomes nonmonotonic that modest coadsorbate stabilization lowers $E_a$ by ~0.20 eV before steric/electrostatic crowding reverses the gain (**Figure 2e**). In contrast, *NO$_2$ cleavage retains an intermediate coverage optimum (down to 0.18 eV near ~2/5 ML; **Figure 2f**). Therefore, coverages that are favorable for mid-sequence *NO$_2$ dissociation steps are not the same as those that favor early N–O scission in *NO$_3$ scission.

Nitrite (NO$_2$⁻) management governs selectivity under the same rules. The formation of *NO$_2$ is exergonic (−0.06 → −0.64 eV across $\theta_{NO3}$), and its $E_a$ is minimized at intermediate $\theta_{NO3}$, so transient *NO$_2$ build-up—and possible NO$_2$⁻ release—is expected (**Figure 2c**). As *NO and *NH$_2$ accumulate, *NO$_2$ adsorption weakens, opening a kinetic escape that erodes NH$_3$ selectivity (**Figure S16**). Maintaining intermediate overall coverage—or modestly strengthening *NO$_2$ binding without overshoot—remains nitrite in the surface cycle and sustains high Faradaic efficiency to NH$_3$.

Intermediate, potential-set coverage maximizes turnover. High coverage pushes different steps in opposite directions: crowding destabilizes *NO$_3$ adsorption and increases the first N–O scission barrier, yet the same coadsorbates can preorganize proton relays and lower key PCET activation barriers. The operating optimum therefore occurs at an intermediate coverage set by the applied potential, where coadsorbate-assisted PCET is preserved but a steric/electrostatic wall to early N–O cleavage has not formed. This step-by-step balance—penalizing *NO$_3$ adsorption/cleavage while aiding mid-sequence hydrogenation—naturally yields the volcano-type activity profile as an ensemble average over coverage-dependent energetics (**Figure 2a** and **2f**).

The same qualitative trends hold under *NO precoverage, with only modest shifts in absolute values (**Supplementary Tables 7--10**). Early *NO$_3$ scission still favors opening, undercoordinated pockets that are penalized on crowded terraces. *NO preadsorption slightly increases the $E_a$ via dipole–dipole repulsion and steric screening but does not invert site ordering. Likewise, *NO hydrogenation toward *NHO remains most sensitive to local crowding, in which densely packed *NO islands suppress proton access and destabilize the PCET transition state. Therefore, preformed *NO adlayers may slightly enhance the potential response but do not alter which microenvironments are intrinsically most productive.

**Coverage-aware Microkinetic Analysis of NO₃RR on Cu(111)**

Even at steady state, the Cu(111) surface is intrinsically fluxional: intermediates adsorb, react, and desorb on the timescale of turnover. To capture these *operando* dynamics, we coupled our constant-potential microkinetic solutions to fresh, MLP-accelerated global structure searches at each applied potential following Stage II of our framework. For every potential between –0.10 and –1.00 V, we enumerated all 450 configurations within 0.50 eV of the global minimum, thereby mapping both the ground state and a spectrum of kinetically accessible metastable states. By explicitly accounting for hundreds of *operando* accessible sites, coverage-explicit microkinetics at constant potential capture the catalytic behavior that the mean-field models miss.[34-36]

Using potential-resolved GC-DFT energetics as the input (**Supporting Methodology** and **Figure S15**), the microkinetic model reproduces a pronounced activity volcano (**Figure 3a**), in which the site-normalized $NH_3$ TOF peaks at 0.015 s$^{-1}$ at −0.70 V and decreases on either side. For $U > −0.70$ V, multiple elementary steps become endergonic, *$NO_2$ accumulates, and partial *$NO_2$ desorption erodes $NH_3$ selectivity. However, for $U < −0.70$ V, stronger intermediate adsorption does not translate into higher rates because the HER siphons electrons and surface *H away, inhibiting nitrate electroreduction. Across −1.00 V to −0.10 V, the modeled Faradaic efficiency for $NH_3$ remains >99% (**Figure 3b**).

Transient species profiles at −0.70 V (**Figure 3c**) reveal the coupling between crowding and intermediate management. $NO_3^-$ is consumed monotonically, whereas $NO_2^-$ exhibits a transient spike: it is generated by *$NO_3$ dissociation, can desorb to solution, and—if retained—readsorbs for further reduction (**Figure S16**). As *NO and *NH$_2$ build up, *NO$_2$ desorption weakens, opening a kinetic window for NO$_2^-$ escape. These dynamics imply two practical levers: accelerate the *NO$_3$ → *NO$_2$ → *NO sequence to minimize desorption opportunities by modestly strengthening *NO$_2$ binding—without overstabilization—to keep it on the surface for completion to NH$_3$.

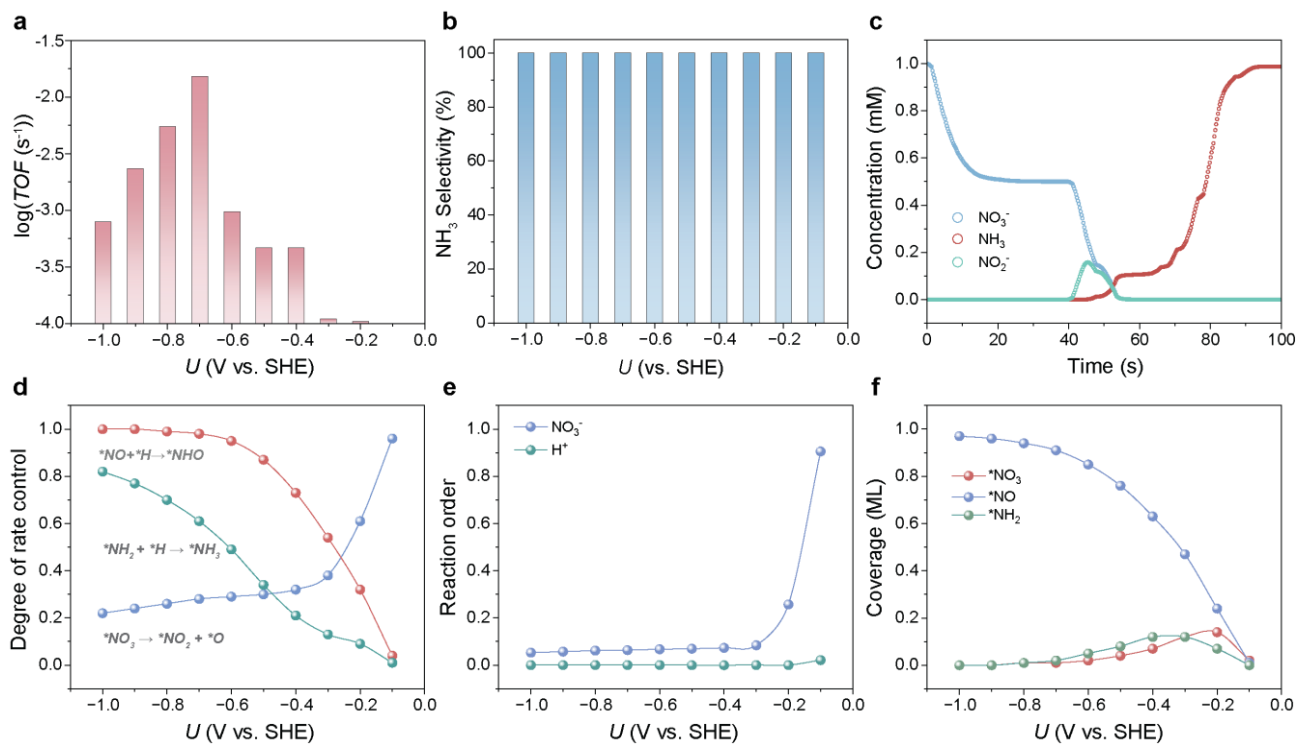

**Figure 3. Coverage-dependent GC–DFT–MKM analysis of the NO$_3$RR. (a)** Volcano-type relationship between the NH$_3$ turnover frequency (TOF) and applied potential. **(b)** NH$_3$ selectivity with NO$_2^-$ as a byproduct across the potential window from –1.00 to –0.10 V (V vs. SHE). **(c)** Time-resolved evolution of NO$_3^-$ (blue), NO$_2^-$ (green), and NH$_3$ (red) concentrations during the reaction at -0.70 V. **(d)** Degree of rate control (DRC) for key steps as a function of the applied potential. **(e)** Rate order analysis with respect to NO$_3^-$ (blue) and H$^+$ (green). **(f)** Surface coverage of the most abundant intermediates—*NO$_3$ (red), *NO (blue), *NH$_2$ (green)—at the steady state.

DRC analysis (**Figure 3d**) revealed a potential-dependent shift in the kinetic bottleneck. Under strongly reducing conditions ($U < -0.70$ V), the first hydrogenation of *NO toward *NHO controls the rate, and lowering this $E_a$ results in the greatest gains at high current density. However, as the applied potential is relaxed toward −0.30 V, rate control transfers downstream to the PCET step *NH$_2$ hydrogenation toward *NH$_3$. At mildly cathodic to near-neutral potentials, the initial *NO$_3$ dissociation again becomes limiting. The apparent reaction orders mirror this picture (**Figure 3e**): the order in [$NO_3^-$] increases steadily with more positive potentials, whereas the order in [$H^+$] remains near zero across the window on Cu(111).

Steady-state coverage rationalizes these shifts (**Figure 3f**). At $U < -0.70$ V, *$NO_3$ dissociation becomes easier, and both *NO hydrogenation and *$NH_2$ hydrogenation are less sensitive to the applied potential. The Cu(111) surface approaches *NO-rich monolayer positioning, aligning with sluggish *NO hydrogenation toward *NHO as the dominant kinetic penalty. As the voltage becomes less negative ($U > -0.70$ V), the combined effect of surface coverage (~0.50 ML) and applied potential on the activation energy ($E_a$) of the elementary steps shifts the rate-determining step from *NO hydrogenation to *$NO_3$ dissociation at −0.30 V (**Figure 5c**).

Over time, this gradually becomes DRC. Progressing to less negative potentials dilutes *NO and admits modest *$NO_3$ and *$NH_2$ populations; by −0.20 V, all three intermediates are sparse, reflecting weak *$NO_3$ adsorption and generally slow downstream steps. Together, the potential and coverage codetermine both the thermodynamics (adsorption/desorption equilibria) and the kinetic hierarchy (barrier ordering). Maintaining an intermediate, "Goldilocks" coverage—set by the applied potential—is essential to maximize activity while preserving near-quantitative $NH_3$ selectivity.

**Dynamic surface ensembles govern activity and selectivity**

Under *operando* applied potential, the Cu(111) interface evolves into a dynamic ensemble of motifs, whose composition and geometry shift with potential. At −0.20 V, the global-minimum (GM) state features an ordered mosaic of *$NO_3$, *NO, and *$NH_2$ in compact mixed islands, with *$NO_3$ stabilizing domain interiors and *NO and *$NH_2$ interlacing along close-packed rows (**Figure 4a**). At −0.70 V, the surface becomes NO-rich ($\theta_{NO} = 0.91$ ML), with traces of *$NH_2$ (**Figure 4b**). The applied potential also influences the width of the *operando* ensemble. At −0.90 to −0.80 V, the landscape is *NO-dominated (GM = 74%), whereas *$NH_2$ introduces a dilute mosaic at −0.70 V (**Figure 4c** and **Figure S18f**). At −0.50 V, *$NO_3$ yields a ternary adlayer and flattens the landscape (GM = 37%) (**Figure S18d**); refunneling follows at −0.40 V (GM = 52%), and dominance is restored by −0.30 V

(GM = 86%; **Figure 4d**).

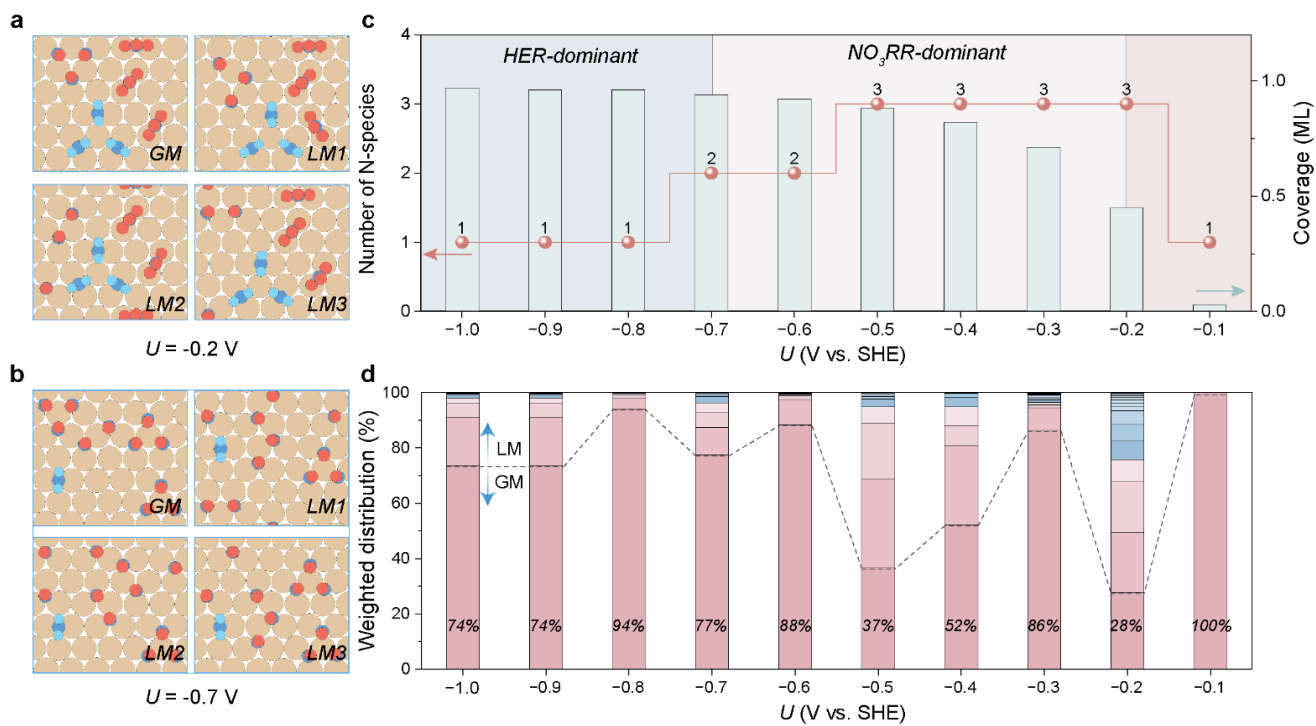

**Figure 4. Dynamic evolution of surface states under varying electrochemical potentials and their corresponding Boltzmann distributions. (a-b)** Representative top views of the global minimum (GM) and local minimum (LM1–LM3) surface states on Cu(111) at **(a)** -0.20 V and **(b)** -0.70 V, respectively. Color code: Cu (yellow), O (red), N (dark blue), H (light blue). The sequence of LM1 to LM3 corresponds to increasingly high-energy metastable states. **(c)** Number of N-containing species at various applied potentials (left y-axis), and the coverage of all N-containing species on 6 × 6 supercell of Cu(111) surface (right y-axis). **(d)** Boltzmann distributions of different surface states across the electrochemical potential window from -0.10 V to -1.00 V. The dashed line represents the boundary between the optimal species and other species. The Boltzmann proportions for each GM are also indicated on the bars. For each potential, the stacked bar chart uses different colors to represent the various surface states, with the most stable structure at the bottom. As the relative energy increases, the states are arranged from bottom to top.

Under reaction conditions, Cu(111) expresses not one "average" site but rather a distribution of microenvironments that repopulates with potential. Clustering >450 potential-resolved structures yields 34 recurrent motifs (**Figure 5d**); at each potential, we rank the four most prevalent microenvironments (Sites 1–4) to compare their intrinsic kinetics at a fixed applied potential. These labels track the local environment (crowding, coordination, coadsorbates), not a rigid atomic pattern— so we can ask, step by step, which sites are fast or slow and why.

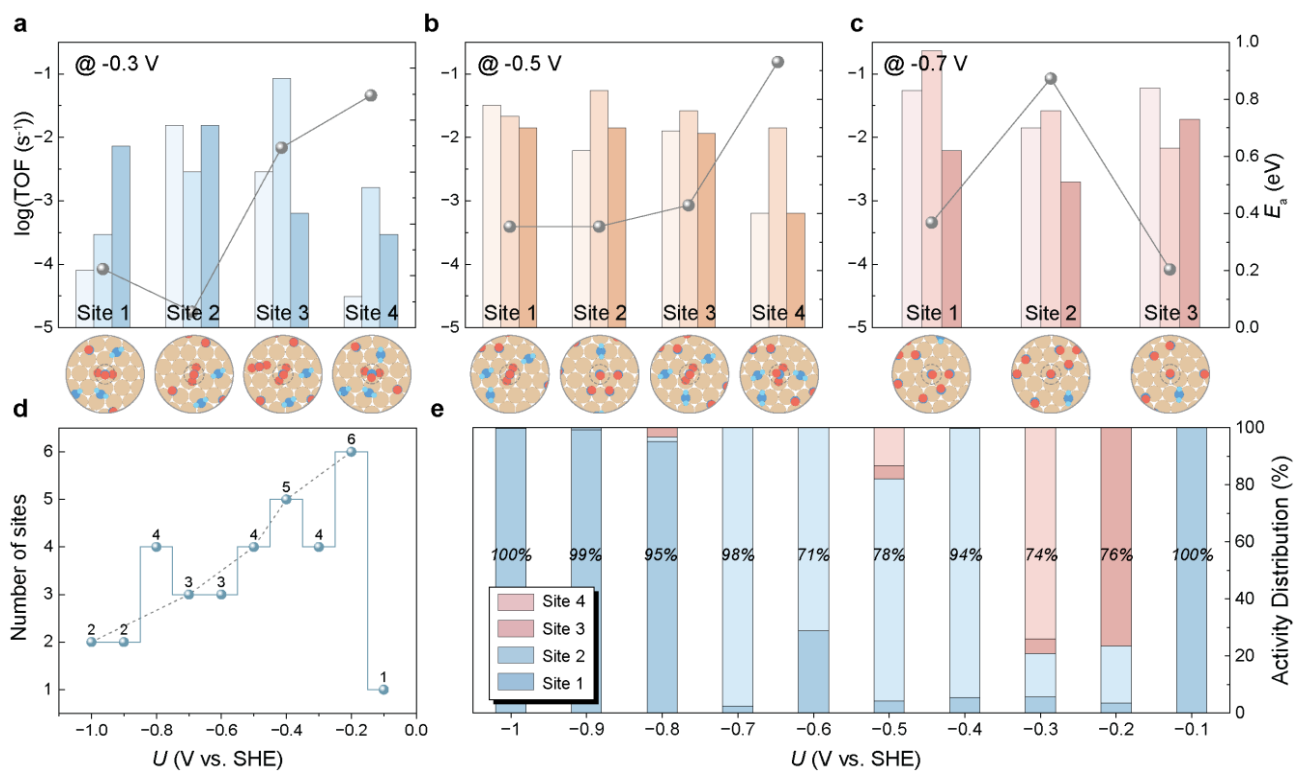

**Figure 5. Active site heterogeneity–induced structure–reactivity relationships and activity contributions of each distinct site motif under varying electrochemical potentials. (a-c)** Activation barriers ($E_a$, column) for DRC—*$NO_3$ dissociation, *NO hydrogenation, and *$NH_2$ hydrogenation—across representative site types and calculated turnover frequencies (TOFs, gray line) for each site (site-specific kinetics) at -0.3 V, -0.5 V, and -0.7 V (vs. SHE), respectively, from left to right. **(d)** Number of distinct active sites accessed at varying applied potentials from -1.00 V to -0.10 V. **(e)** Activity contribution of each site motif under varying electrochemical potentials, with site 1 (dark blue) being the most prevalent.

At −0.30, −0.50, and −0.70 V, the operando ensemble contains 4, 4, and 3 representative sites, respectively (**Figure 5a–d**). We quantify each site's contribution by combining its intrinsic per-site TOF with its population. At −0.30 V, the low-population Site 4 dominates the overall $NO_3RR$ flux because it exhibits the lowest activation barriers, outperforming more prevalent terrace sites. By contrast, at −0.50 and −0.70 V, the highly populated Site 1 and Site 2 provide the largest share of $NH_3$ production despite the high intrinsic activity of Site 4 (**Figure 5**). Below −0.70 V, Sites 1 and 2 still account for most of the production because their surface populations are large (**Figure 5e**). Thus, the macroscopic NO$_3$RR rate is controlled by the product of intrinsic kinetics and site population—neither factor alone is predictive.

Zooming in at −0.70 V (the activity maximum), the interface expresses three sites: Site 1 (crowded terrace), Site 2 (stepped, lightly covered), and Site 3 (highly crowded terrace). At this potential, early N–O scission in *NO$_3$ is fastest at sparse, stepped sites (**Figure 5c**). The *NO$_3$ dissociation barrier is ~0.84 eV at Site 1 and Site 3 but decreases to 0.70 eV at Site 2 (**Figure S19f**). This trend holds across potentials: open, undercoordinated Cu pockets accelerate O–N cleavage, whereas confined terraces slow it through steric/electrostatic blockade. Therefore, Site 2 provides the most favorable gateway for the rate-setting cleavage step.

The mid-sequence PCET (*NO hydrogenation toward *NHO) reversed this ordering. On congested terraces, local dipoles and crowding destabilize the PCET transition state. In mixtures, more open motifs, short proton-relay contacts and access to an undercoordinated Cu atom lower the barrier. At −0.70 V, *NO hydrogenation is slowest at Site 1 (dense *NO trimer island; $E_a$ = 0.97 eV), intermediate at Site 2 (mixed 2NO/NH$_2$ triangle adjacent to a *NO trimer island; $E_a$ = 0.76 eV), and fastest at Site 3 (more open terrace; $E_a$ = 0.63 eV). In contrast, the late *NH$_2$ hydrogenation to *NH$_3$ step is comparatively site insensitive (~0.60 eV), which is consistent with a compact transition state weakly coupled to lateral fields.

Overall turnover reflects the balance between scission and downstream hydrogenation (**Figure S19**). At −0.70 V, Site 1 formed stable *NO trimer islands, whereas dipole screening and site blocking limit activity to ~10$^{-3}$ s$^{-1}$. Site 3, with two *NO and one *NH$_2$ far apart, lacks cooperative stabilization and efficient proton delivery, resulting in ~10$^{-4}$ s$^{-1}$. In contrast, Site 2—adjacent to a trimer but hosting a mixed 2NO/NH$_2$ microensemble—preserves an open Cu pocket and short NH$_2$···O(NO) relays, reaching ~10$^{-1}$ s$^{-1}$, two orders of magnitude higher than Sites 1 and 3. Therefore, productivity is not the same as prevalence, as minority, site-2-like motifs can set the overall rate even when structurally dominant environments contribute little per site.

The two design rules are generalized from −1.00 to −0.10 V (**Figure 5a-c** and **Figure S19**). (i) *NO trimers are intrinsically slow: packing three *NO on 3–4 Cu atoms overstabilizes reactants, screens dipoles, corrals proton donors, and flattens the potential response (small d$E_a$/d$U$), rendering both *NO hydrogenation and *NO$_3$ dissociation sluggish. (ii) Mixed 2NO/2NH$_2$ quadrilaterals are intrinsically fast in that they disrupt *NO clustering, retain an undercoordinated Cu pocket, and create short NH$_2$···O(NO) proton relays that modestly destabilize *NO and preorganize the PCET transition state. As the potential is driven more negatively, the highest site-specific TOFs concentrate on these mixed motifs, whereas trimer-rich islands remain kinetically inert despite stronger driving.

Motif contributions evolve with potential (**Figure 5e**). At −0.90 V, a single hollow-bound *NO motif (often Site 1) accounts for ~85% of NH$_3$ formation; by −0.50 V, its share decreases to ~52% as the lower population yet intrinsically faster motifs (notably 2NO/2NH$_2$-like) take over. Practically, rather than stabilizing a single "privileged" geometry, tuning the potential, composition, electrolyte, and strain to increase the expression of 2NO/2NH$_2$-assisted, undercoordinated environments offers a route to maximize nitrate electroreduction on Cu(111).

**Localized direct versus surface-mediated adsorbate interactions**

To understand why ostensibly similar motifs turn over at different rates, we disentangle local, direct adsorbate–adsorbate couplings (sterics, dipoles, proton-relay H-bonds) from surface-mediated responses captured by three orthogonal descriptors (**Figure 6a**): (i) the site-resolved excess charge on Cu ($\Delta q_{Cu}$), (ii) the surface Cu $d$-band center, and (iii) lattice distortion in the top Cu layer. Together, these factors provide a compact link from the applied potential → electronic response → structure → reactivity.

$\Delta q_{Cu}$ is the primary kinetic dial. As the applied potential is driven cathodically, the local electron density increases $\Delta q_{Cu}$ (~1.2 → ~5.0 $e$ per (6×6) slab), sculpting a heterogeneous $\Delta q_{Cu}$ landscape (**Figure 6c**). Mapping activation barriers for the rate-determining step onto $\Delta q_{Cu}$ collapses >150 transition states into two near-linear families (**Figure 6e** and **6f**): in electron-rich pockets (more negative $\Delta q_{Cu}$), the *NO$_3$ dissociation barrier decreases by up to ~0.60 eV, whereas as $\Delta q_{Cu}$ approaches neutrality, *NH$_2$ hydrogenation toward the *NH$_3$ PCET barrier decreases by ~0.30 eV. Therefore, early N–O cleavage prefers $\Delta q_{Cu}$ <0, whereas late N–H formation prefers $\Delta q_{Cu}$≈0—elevating local charge redistribution to a unifying kinetic descriptor.

The *d*-band and the lattice transmit the electronic landscape set by $\Delta q_{Cu}$. As the interfacial charge accumulates, the surface *d*-band center shifts downward from −2.31 to −2.50 eV (**Figure 6c**), weakening σ-donation to oxygenates and alleviating *$NO_3$/*NO overbinding. In parallel, the top Cu layer becomes more rumpled (RMS out-of-plane ~2.4→~2.8 Å; **Figure 6b**), creating low-coordination protrusions that colocalize with electron-rich regions (**Figure 6d**). These protrusions disrupt *NO clustering, maintain undercoordinated Cu pockets, and lower barriers for *NO hydrogenation and downstream *$NH_x$ formation. In short, *d*-band shifts set the stage, and lattice distortion amplifies it, but $\Delta q_{Cu}$ plays an important role.

Once $\Delta q_{Cu}$, the *d*-band, and the lattice select the "right" patches, direct adsorbate–adsorbate couplings set the pace. Crowding and dipole screening in *NO trimer islands slow *NO hydrogenation and *$NO_3$ scission, whereas short *$NH_2$···O(NO) contacts act as proton relays that preorganize the PCET transition state. The mixed 2NO/2$NH_2$ quadrilateral is the winning microensemble that tunes $\Delta q_{Cu}$ to an intermediate regime, breaks *NO clusters, installs proton-relay geometries, and consistently delivers the highest turnover across potentials.

Design implication — Program the ensemble by writing $\Delta q_{Cu}$. Practical levers—electrolyte cations, dilute alloying, tensile strain/defects, and pulsed bias—operate by rewriting the $\Delta q_{Cu}$ map, with *d*-band shifts and lattice rumpling as coresponses. The goal is to generate charge-rich protrusions to accelerate early *$NO_3$ cleavage while preserving charge-lean, moderately crowded pockets for late PCET. This is ensemble programming by steering motif populations and local charges to express the rare—but intrinsically fastest—microenvironments at the intended operating applied potentials.

The applied potential sets the interfacial Fermi level; the Fermi level writes the $\Delta q_{Cu}$ landscape; $\Delta q_{Cu}$, in turn, sculpts both the lattice distortion and the surface *d*-band position. This coupled electronic–geometric response redistributes adsorbates, selectively activating the microenvironments that carry the catalytic flux. Restructuring is therefore the operative environment—not a byproduct—linking charge, structure, and coverage to control activity and selectivity.[37-42] Accordingly, tuning $\Delta q_{Cu}$ via electrolyte cations,[43, 44] dilute alloying,[45-47] strain/defects,[48-50] or pulsed bias[22, 51-53] offers concrete handles to repopulate the most productive motifs—targeting charge-rich sites for N–O scission while maintaining charge-lean pockets for N–H formation—and thereby coordinating speciation, structure, and kinetics under operando conditions (**Supporting Note 7**).

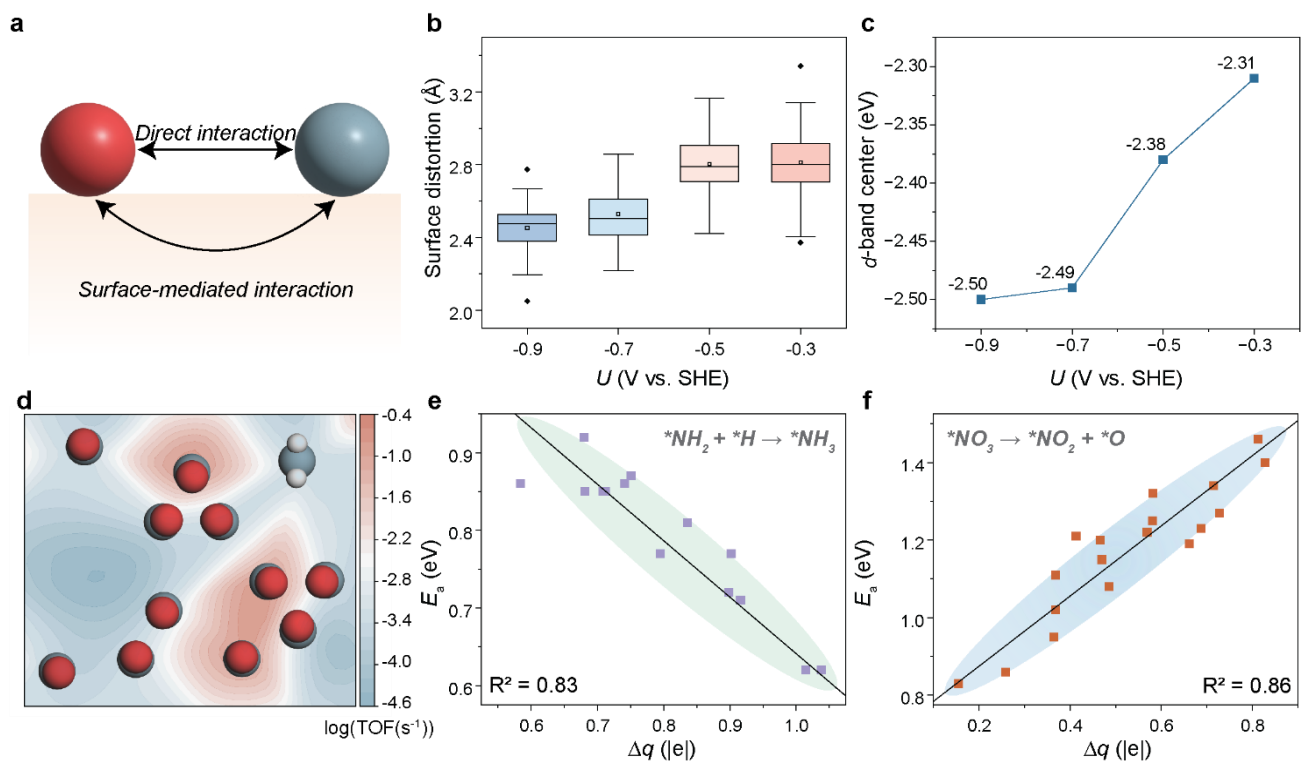

**Figure 6. Localized direct and surface-mediated interactions between adsorbents and adsorbents. (a)** Schematic of adsorbate–adsorbate interactions and adsorbate–catalyst interactions. **(b)** Surface distortion of the Cu(111) top layer atoms under the corresponding surface states. **(c)** $D$-band center and surface charge transfer of Cu(111) top layer Cu atoms. **(d)** Electronic structure of the most active site motif (charge difference). **(e)** Linear correlation between *$NO_3$ dissociation barriers and site-specific charge perturbations ($\Delta q$) across three different surface environments. **(f)** Linear correlation between *$NH_2$ dissociation barriers and site-specific charge perturbations ($\Delta q$) across three different surface environments.

## Discussion

We recast nitrate electroreduction on Cu(111) as an ensemble-controlled process rather than a single-site reaction. By explicitly enumerating and population-weighting 33 recurrent adsorbate microenvironments at constant potential, the framework reproduces the activity volcano and the abrupt loss of NH$_3$ selectivity at more anodic bias—without fitted parameters. A central implication is the decoupling of thermodynamic prevalence from catalytic productivity: the population-dominant ground state at very negative potentials contributes little to flux, whereas rarer motifs near −0.70 V carry a disproportionate share of turnover. This population–productivity mismatch clarifies why mean–field models, which collapse heterogeneity into an "average" site, systematically mispredict rates and selectivity.

The ensemble view provides a coherent mechanistic picture and reconceives "coverage." The applied potential sets the interfacial Fermi level and thereby writes a heterogeneous landscape of site-resolved excess charge on Cu ($\Delta q_{Cu}$). Across >150 activation barriers—from early N–O scission to late proton-coupled electron transfer—activation energies collapse into a single linear relationship with $\Delta q_{Cu}$, increasing the interfacial charge redistribution to a unifying kinetic descriptor. The volcano maximum at −0.70 V emerges when $\Delta q_{Cu}$ is tuned to an intermediate value realized by a specific 2NO/2NH$_2$ quadrilateral that simultaneously lowers the barriers for N–O cleavage and N–H formation. Thus, the macroscopic activity profile reflects potential-driven repopulation of microenvironments with distinct $\Delta q_{Cu}$ values and lateral interactions, not a simple Sabatier balance on a uniform surface.

The predicted kinetics are in accordance with the *operando* signatures of key intermediates and reaction progression. The most abundant surface species—*NO$_3$, *NO, and *NH$_2$[21, 22]—form a coadsorbative network that governs activity and selectivity.[53] At potentials that are milder than the peak, slow *NO$_3$ activation limits turnover; at more negative potentials, the HER diverts electrons and *H despite stronger intermediate adsorption.[54, 55] Transient NO$_2^-$ accumulation arises naturally from crowding and diminishes as *NO/*NH$_2$ builds, creating both a desorption pathway and an opportunity for readsorption and subsequent reduction. Selectivity follows this logic: at more positive potentials, NO$_2^-$ appears as a byproduct,[56] whereas shifting to more negative potentials drives NH$_3$ selectivity toward ~100%—tempered by increasing HER competition.[54] Surfaces engineered to retain nitrite just long enough for further reduction can reconcile this kinetic mismatch.[57-59]

These insights translate into actionable levers. Maintaining "Goldilocks" coverage—via electrolyte formulation and mass transport—preserves beneficial coadsorbate stabilization while avoiding steric/electrostatic blockade. Pulsed-bias protocols can transiently favor *NO$_3$ activation when $\Delta G_{ads}$(*NO$_3$) < 0, followed by milder conditions that relieve crowding and complete hydrogenation/desorption. Electronic and structural tuning—dilute alloying with electron-donating elements, interfacial cation engineering, or tensile strain—can embed charge-rich protrusions that accelerate early N–O scission adjacent to charge-lean, less crowded regions that expedite late N–H formation. Each lever operates by steering $\Delta q_{Cu}$ and, consequently, the population of the most productive motifs.

Two caveats set limits and opportunities. First, we analyze ideal Cu(111) terraces and do not explicitly include steps, defects, or oxide remnants, which would widen the set of accessible motifs. These features can be added within the same population-based framework. Second, solvent dynamics and ion-specific effects beyond our grand canonical treatment may further tune $\Delta q_{Cu}$ and lateral coupling. As improved continuum or hybrid explicit–implicit models become practical, they can be integrated. Despite these limits, the main point is robust: on Cu(111), the performance arises from a voltage-selected distribution of local environments governed by interfacial charge.

More broadly, we replace the scalar "coverage" with a population of local microenvironments, each with its own kinetics. This ensemble-kinetics view is transferable to high-coverage electrocatalysis. By combining constant-potential energetics, coverage-explicit microkinetics, and global structure sampling, it introduces a quantitative control knob—ensemble population weighting—tunable by composition, electrolyte, strain, and dynamic bias. Shifting the question from "what is the active site?" to "which ensemble dominates at a given driving force?" enables rational, population-level programming of catalytic interfaces beyond nitrate-to-ammonia.

## Conclusions

We redefine "coverage" as a population of local microenvironmental motifs with distinct kinetics rather than a single scalar. Nitrate electroreduction on Cu(111) is governed by dynamic, potential-selected ensembles rather than a single static site. A machine-learning-accelerated, constant-potential multiscale workflow identifies 33 recurrent motifs whose populations and intrinsic activities coevolve with bias. The framework explains the pronounced activity volcano peaking at −0.70 V (vs. SHE), with a site-normalized turnover frequency of ~0.015 s$^{-1}$ and near-quantitative NH$_3$ selectivity, and clarifies its deterioration at more anodic or cathodic potentials. The controlling physics reduce to a simple descriptor: the excess charge on the reactive Cu atom ($\Delta q_{Cu}$), which linearly correlates with barriers across diverse local environments. This charge-centric view decouples stability from productivity, rationalizes operando fluxionality, and provides concrete levers—electronic/structural heterogeneity, electrolyte cations, and potential pulsing—for "ensemble programming". Recognizing and incorporating such dynamic interfacial conditions is essential for predictive kinetics and design. The strategy demonstrated here—explicit motif enumeration, constant-potential energetics, and coverage-resolved microkinetics—offers a transferable blueprint for high-coverage electrocatalysis beyond the NO$_3$RR and a path toward catalysts that exploit, rather than suppress, their intrinsic dynamism.

## Computational Details

**NNP training and global minimum structure search**

The neural network potential training process is based on the REANN package.[11] To collect dataset structures covering the chemical space we focus on, we sampled three main types of atomic structures: a Cu (111) bare surface and a Cu (111) surface covered with different types of intermediates to fully capture the lateral adsorbate interactions between reaction intermediates. The dataset used for neural network potential training and prediction is a collection of diverse sets of calculations corresponding to all nitrogen-containing intermediates (*NO$_3$, *NO$_2$, *NO, *NHO, *NOH, *NHOH, *NH$_2$OH, *N, *NH, *NH$_2$, *NH$_3$) on the Cu(111) surface with various coverages and under mixed multi-intermediate coverages. The genetic algorithm is a population-based stochastic procedure employing the survival of the fittest principle in our previous work. This technique has been applied to determine the surface structures at given compositions of surface intermediates.

**Model Setup and DFT Calculations**

The Cu(111) surface is modeled by a 6 × 6 supercell of Cu(111) termination with a cell dimension of 15.336 Å × 15.336 Å (constructed with an experimental lattice parameter). The bottom two layers of the slab are constrained as the bulk region, and everything else is allowed to relax as the interface region. A vacuum slab of 15 Å thickness is added in the z direction to avoid spurious interactions between periodic images. The calculational structure was visualized via OVITO software.[10]

Spin-polarized DFT calculations were performed via the projector-augmented wave (PAW) method and the Perdew–Burke–Ernzerhof (PBE) generalized gradient approximation for exchange-correction functional with a 400 eV plane wave energy cutoff in the Vienna ab initio simulation package.[60-62] The Brillouin zone sampling was restricted to the Γ point with a 1×1×1 mesh and a Monkhorst-Pack 2×2×1 mesh.[63] The convergence criterion for the electronic self-consistent loop was set to $10^{-4}$ eV. Transition states were identified with the automated nudged elastic band method and dynamic nudged elastic band method (Dy-NEB) to produce a good initial guess via the image-dependent pair potential (IDPP) surface method to locate transition states with REANN as initial structures and identified with a force tolerance of 0.05 eV/Å via multiple methods, including the climbing image nudged elastic band (CI-NEB) method and improved dimer method.[64-67] Vibrational mode analysis was also conducted to validate the identified transition states. Bader charge analysis, as implemented by the Henkelman group, was also performed.[68]

**Data availability**

The data generated in this study are provided in the article and the Supplementary Information files.

## Acknowledgments

This work was supported by the Key Technologies R&D Program of China (2021YFA1502804), the National Natural Science Foundation of China (22172150, 22221003 and 22222306), and the Innovation Program for Quantum Science and Technology (2021ZD0303302). We thank the robotic

## Competing interests

The authors declare that they have no competing interests.

**TOC**

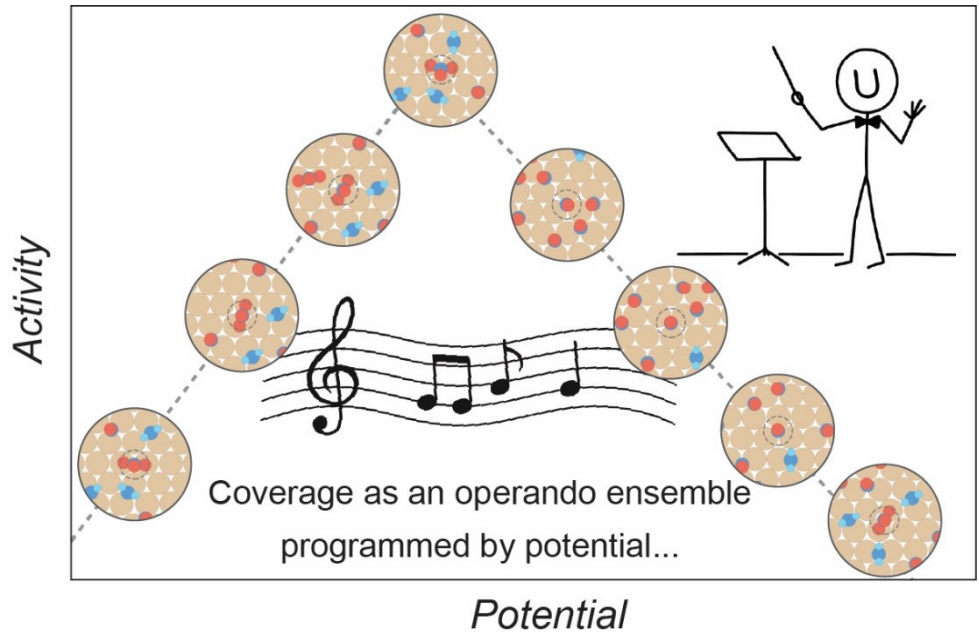